\documentclass[amsmath,amssymb,aps,graphicx,twocolumn,reprint,linenumbers]{revtex4-2}
\usepackage{graphicx}
\usepackage{physics}
\usepackage{color}
\usepackage{mathrsfs}
\usepackage[caption=false]{subfig}

\begin{document}
\newcommand{\dif}{\mathrm{d}}

\nolinenumbers

\title{An ultra-high gain single-photon transistor in the microwave regime}

\author{Zhiling Wang$^{1}$}
\altaffiliation{These three authors contributed equally to this work.}

\author{Zenghui Bao$^{1}$}
\altaffiliation{These three authors contributed equally to this work.}

\author{Yan Li$^{1}$}
\altaffiliation{These three authors contributed equally to this work.}

\author{Yukai Wu$^{1,2}$}

\author{Weizhou Cai$^{1}$}

\author{Weiting Wang$^{1}$}

\author{Xiyue Han$^{1}$}

\author{Jiahui Wang$^{1}$}

\author{Yipu Song$^{1,2}$}

\author{Luyan Sun$^{1,2}$}

\author{Hongyi Zhang$^{1,2}$}
\email{hyzhang2016@tsinghua.edu.cn}

\author{Luming Duan$^{1,2}$}
\email{lmduan@tsinghua.edu.cn}

\affiliation{$^{1}$Center for Quantum Information, Institute for Interdisciplinary Information Sciences, Tsinghua University, Beijing 100084, PR China}
\affiliation{$^{2}$Hefei National Laboratory, Hefei 230088, PR China}

\date{\today}

\begin{abstract}

A photonic transistor that can switch or amplify an optical signal with a single gate photon requires strong non-linear interaction at the single-photon level.
Circuit quantum electrodynamics provides great flexibility to generate such an interaction, and thus could serve as an effective platform to realize a high-performance single-photon transistor.
Here we demonstrate such a photonic transistor in the microwave regime. Our device consists of two microwave cavities dispersively coupled to a superconducting qubit. A single gate photon imprints a phase shift on the qubit state through one cavity, and further shifts the resonance frequency of the other cavity. In this way, we realize a gain of the transistor up to 53.4 dB, with an extinction ratio better than 20 dB. Our device outperforms previous devices in the optical regime by several orders in terms of optical gain, which indicates a great potential for application in the field of microwave quantum photonics and quantum information processing.

\end{abstract}

\maketitle 
\nolinenumbers

\section{Introduction}
The fact that photons hardly interact with each other makes them excellent carriers of information for long-distance quantum communication~\cite{Cirac1997}, but on the other hand, challenges the realization of all-optical quantum operation which is necessary for photonic quantum networks and all-optical quantum devices~\cite{Kimble2008}.
An effective photon-photon interaction can be generated through the strong light-matter interaction by introducing non-linear optical medium, such as atomic ensembles~\cite{Hofferberth2014,Rempe2014} and cavity quantum electrodynamic (QED) systems involving single atoms~\cite{Vladan2013,Dayan2014,Stephan2014,Lukin2014}, single artificial atoms~\cite{Waks2016,Waks2018,Wallraff2008,Imamoglu2012} or exciton-polaritons~\cite{Pavlos2021}. 
Photon-photon interaction at the single-photon level enables the realization of a single-photon transistor (SPT), in analogy to an electronic transistor, that can switch or amplify ``source'' photons controlled by a single ``gate'' photon~\cite{Lukin2007}.
SPT lies at the heart of many important applications including optical computing~\cite{Furusawa2009} and hardware-efficient quantum random access memory (qRAM)~\cite{qram2008,Englund2021}.
In the optical regime, SPT has been realized in some pioneering works~\cite{Vladan2013,Hofferberth2014,Rempe2014,Waks2018}, but with limited performance with the highest gain of less than 100, which is mainly due to the imperfect quantum control or insufficient coupling strength between light and matter.

As an alternative, circuit QED architectures based on the superconducting quantum circuit have been proposed as an appealing platform to realize a SPT in the microwave regime~\cite{Hartmann2013,Anders2016,Stolyarov2020,Anders2014}. Quantum operation between a propagating single microwave photon and superconducting qubit has been well-established~\cite{Nakamura2018,Wallraff2018,Jonathan2020review,Wallraff2021}, benefiting from the strong and well-controlled interaction between them. As a consequence, high-efficient photon-photon interaction with great flexibility is attainable at the single-photon level~\cite{reuer2021realization}. 
Moreover, the conveniently engineered superconducting qubit and high-quality superconducting microwave cavities endow higher-order nonlinear effects~\cite{Oliver2020,Blais2021}, which are currently unreachable in cavity QED systems, potentially facilitating the realization of a SPT with extraordinary performance.

In this work, we have realized a SPT in the microwave regime with unprecedented gain up to 53.4 dB, together with an extinction ratio above 20 dB. Our device relies on a controlled-phase gate between a propagating single microwave photon and a superconducting qubit, which triggers the succeeding switch of the input signal. The gate photon can be well-preserved after the switch operation, which enables us to confirm the single-photon nature of the switching process.
Our work demonstrates a great possibility of photonic quantum information processing in the microwave regime~\cite{Jonathan2020review,GU2017}, such as coherent state manipulation~\cite{Ritsch1997} and the realization of hardware-efficient qRAM~\cite{qram2008,Englund2021}.

\begin{figure*}[!tbp]
\centering
\includegraphics[width=0.9\linewidth]{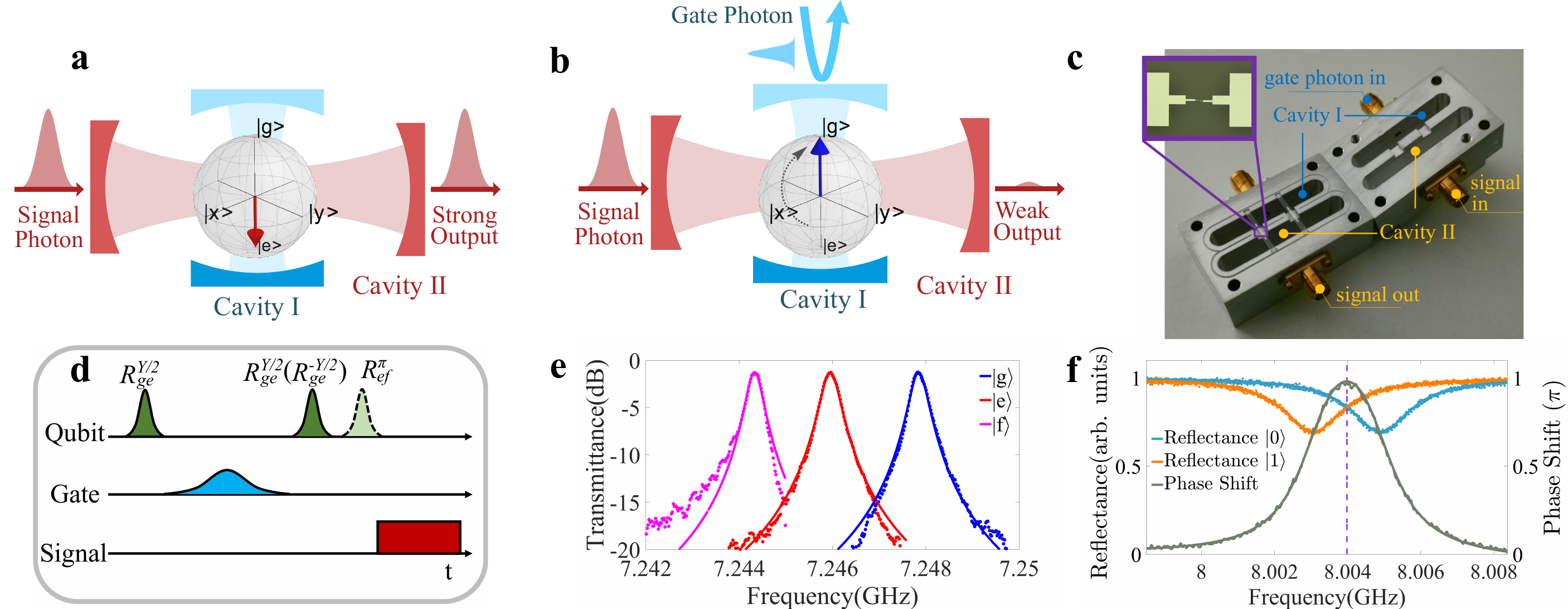}
\caption{\textbf{A single-photon transistor.} Our microwave single-photon transistor consists of a superconducting qubit dispersively coupled with two superconducting cavities. Cavity I is used to couple the gate photon with the qubit which in turn controls the transmission of signal photons through Cavity II.  \textbf{a}, When the gate pulse contains no photon, the qubit state would be flipped to either $\ket{e}$ or $\ket{f}$. If the frequency of the signal photons is resonant with the dispersively shifted frequency of cavity II, most of the signal photons would pass through the cavity, and the transistor is ``on''. \textbf{b} When the gate pulse contains a single photon, the qubit would stay at $\ket{g}$ after the pulse sequence. The signal photons are off-resonant with the frequency of cavity II, and thus few signal photons can pass through the cavity. Consequently, the transistor is switched ``off'' by the single gate photon. 
\textbf{c} shows a picture of the single-photon transistor. The inset picture is a micrograph of the superconducting qubit. The qubit transition frequency is $\omega_q/2\pi=5.350$ GHz, and the qubit anharmonicity $E_c/2\pi = 249$ MHz. \textbf{d} shows the pulse sequence for the operation of the single-photon transistor. The transistor can be either operated in qubit $\{\ket{g}$,$\ket{e}\}$ subspace, or in $\{\ket{g}$,$\ket{f}\}$ subspace, depending on whether the last qubit $\pi_{ef}$ pulse is applied. \textbf{e} shows the transmittance of cavity II when the qubit state is in $\ket{g}$, $\ket{e}$ and $\ket{f}$. \textbf{f} shows the qubit state dependent reflectance of cavity I (in orange and light blue), and the resulting phase difference of the qubit-state-dependent reflections (in light green). The dashed purple line indicates the frequency of the gate photons.}
\label{scheme}
\end{figure*}

\section{Results}

\bigskip
\textbf{The single-photon transistor.} An illustration of our single-photon transistor can be found in Fig.~\ref{scheme}\textbf{a} and \textbf{b}. An incoming gate photon pulse controls the transmittance of the signal photons fed to the transistor. If the gate pulse contains no photons, the signal photons could pass through the transistor. If the gate pulse contains one photon, the signal photons could barely travel across the transistor. The gate photon survives after the switch operation. In the experiment, we use two microwave cavities dispersively coupled with a superconducting transmon qubit to separately detect the gate photons (cavity I) and switch the signal photons (cavity II), as shown in Fig.~\ref{scheme}\textbf{c}. Cavity I is designed as a single-side cavity to minimize photon loss during the detection of gate photons, while cavity II is a double-side one to maximize the transmittance of signal photons. The out-coupling rate of cavity I is fine-tuned to meet $\kappa_I =  2\pi\times1.81 $ MHz $\approx 2|\chi_I^{ge}| =2\pi\times1.73 $ MHz, where $\kappa_I$ is the linewidth of cavity I, and $\chi_I^{ge}$ is the qubit excited state $\ket{e}$ induced dispersive shift for cavity I. In this way the reflected single gate photon would have a $\pi$ phase shift conditioned on the excited state of the qubit~\cite{Nakamura2018,wang2021flying}, as indicated in Fig.~\ref{scheme}\textbf{f}. Such a controlled-phase gate between the gate photon and the qubit forms the basis for gate photon detection, which has been successfully demonstrated in previous works~\cite{Nakamura2018,Wallraff2018}.

To operate the single-photon transistor, as shown in Fig.~\ref{scheme}\textbf{d}, the qubit is initially prepared in $(\ket{g}+\ket{e})/\sqrt{2}$. If a gate photon is sent to cavity I, the qubit-photon state would be $(\ket{1}\ket{g}-\ket{1}\ket{e})/\sqrt{2}$, which can be also identified as a conditional qubit phase flip in the presence of the single gate photon~\cite{Duan04cz,Nakamura2018,Wallraff2018,Wallraff2020,wang2021flying}. A following $\pi/2$ gate would drive the qubit to $\ket{g}$, as illustrated in Fig.~\ref{scheme}\textbf{b}. On the other hand, as shown in Fig.~\ref{scheme}\textbf{a}, if there is no gate photon sent to cavity I, the qubit state would be in $\ket{e}$ after the pulse sequence. 
Cavity II is used to switch the signal photons according to the qubit state. The out-coupling rates for the input and the output ports of this cavity are tuned to $\kappa_{II}^{in}=\kappa_{II}^{out}= 2\pi\times0.13$ MHz to achieve a maximized transmittance at the resonance frequency. We further take $2|\chi_{II}^{ge}| = 2\pi\times1.894$ MHz $\gg \kappa_{II} = 2\pi\times0.3$ MHz to guarantee that the qubit state depended cavity resonance shift is much larger than the linewidth of cavity II, as indicated in Fig.~\ref{scheme}\textbf{e}. Consequently, the signal photons can be effectively switched off (on) on the transmission path once the resonance frequency of cavity II is shifted by the gate photon determined qubit state.
We note that it is also possible to operate the transistor within the qubit $\{\ket{g}$,$\ket{f}\}$ subspace by applying a $\pi_{ef}$ pulse after second $\pi_{ge}/2$ pulse. In this case one would have a larger dispersive shift $2|\chi^{gf}_{II}| = 2\pi\times3.518$ MHz but could suffer from stronger decay/dephasing of the $\ket{f}$ state as a compromise.

\begin{figure*}[!tbp]
\centering
\includegraphics[width=0.7\linewidth]{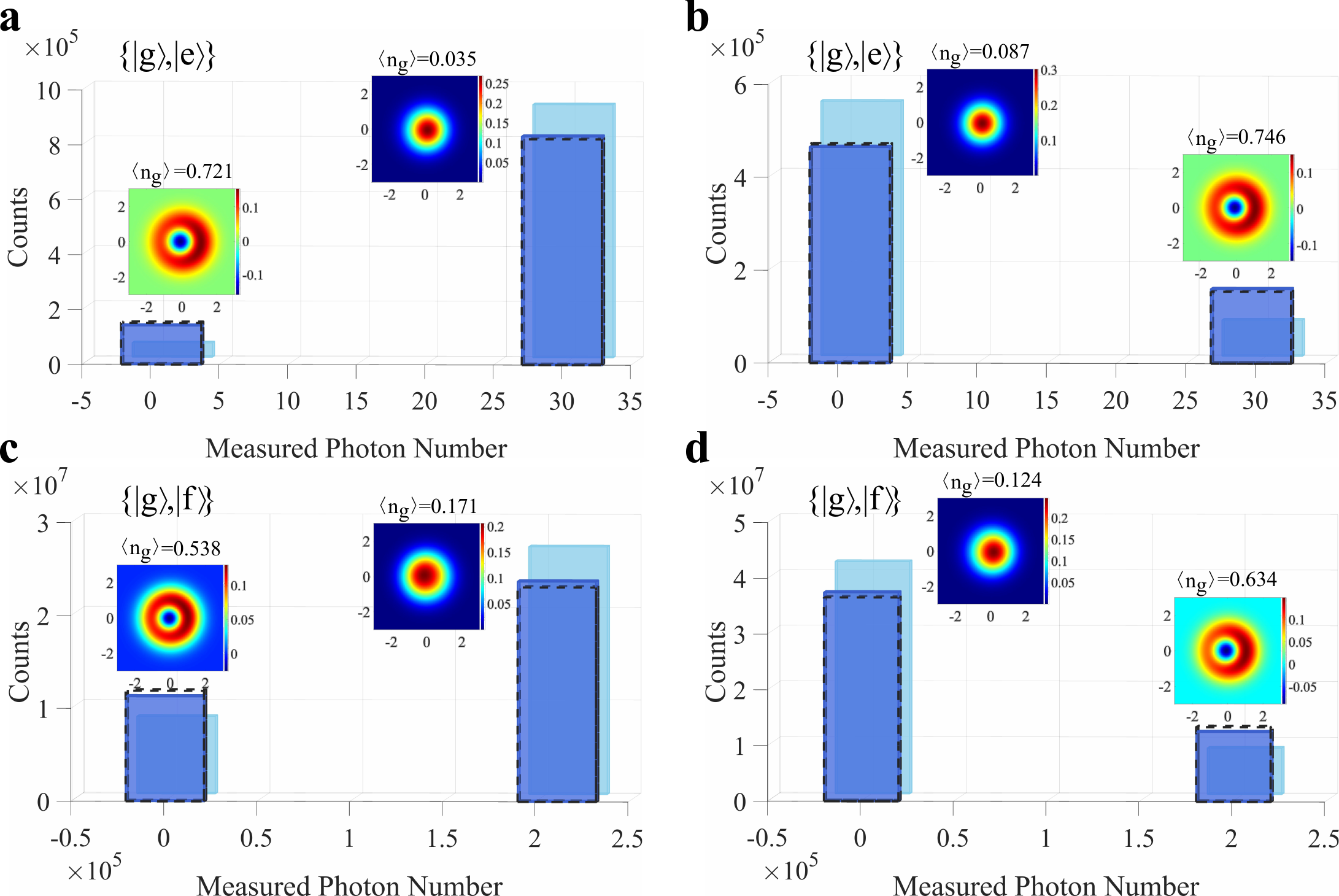}
\caption{\textbf{Single-photon switch.} We use a Gaussian-shaped photon pulse in the weak coherent state with an average photon number $n_g$ =0.18 to gate the transistor. The frequency of the signal photon is aligned to the resonance frequency of cavity II when the qubit is in either $\ket{e}$ or $\ket{f}$. \textbf{a} presents statistical results for the ``on''/``off'' states of the transistor with an average signal photon number $n_s$=37.2 and qubit phase $\theta=0$. The difference between the statistics for the transistor state with (dark blue bars) and without (light blue bars) gate photons demonstrates a successful switch by the randomly arrived single gate photon. The reason that the difference is relatively small is mainly due to the low photon number in the gate pulse, thus the transistor can not be effectively switched off. The dotted bars show the corresponding theoretical results (see Supplementary Note 6 B for details). The x-axis position for the bars gives the average output photon number for each case. The insets are the reconstructed Wigner function of the gate photons reflected from cavity I, conditioned on the weak or strong transmission, respectively. The Wigner function of the gate photons conditioned on a weak transmission resembles a vacuum state, whereas for a strong transmission, the measured Wigner function is similar to a single-photon Fock state. The conditional average photon number is listed. \textbf{b} shows similar experimental results with $n_s$= 37.2 and $\theta=\pi$, where the transistor is working as normally closed. The experimental results when using an ultra-strong signal strength with $n_s$= 2.62$\times10^5$ are shown in \textbf{c} and \textbf{d}, with $\theta=0$ and $\theta=\pi$, correspondingly. Both the gate photon number and the signal photon number are calibrated based on the AC stark shift induced dephasing of the transmon qubit. Details about this calibration can be found in Supplementary Note 2.}
\label{switch}
\end{figure*}

\bigskip
\textbf{The switch effect.} We first characterize the switch effect of the single-photon transistor. A 10 $\mu$s long square pulse with an average photon number $n_s$=37.2 is used as the input signal, whose frequency is aligned to the resonance frequency of cavity II when the qubit is in $\ket{e}$. From Fig.~\ref{scheme}\textbf{a}, the qubit is expected in $\ket{e}$ if there is no gate photon, and thus the transistor is normally open. In the experiment, a Gaussian pulse in the weak coherent state, with a pulse length of 960 ns and an average photon number $n_g=0.18$, is used to simulate a single-photon source for gate control. The transmission of the input signal with and without the gate photons is repeatedly measured to determine the ``on'' or ``off'' state of the transistor in each trial (see Methods for details), as illustrated with the bar plots in Fig.~\ref{switch}\textbf{a}. Compared with the results without gating photons, one could see a clear decrease in the count of ``on'' events, while a correspondingly increased count of the ``off'' events when the transistor is gated, which demonstrates the success of switch operation. The reason that the transistor is not largely switched off is mainly that a large portion of the weak coherent pulse is still in the vacuum state, while for this demonstration we do not have a single microwave photon detector for post-selection as in previous optical experiments~\cite{Waks2018}.

\bigskip
\textbf{Single-photon switch.} To confirm the single-photon switching character, we evaluate the quantum state of the reflected gate photons conditioned on either the ``on'' or ``off'' state of the transistor, which can be determined in a single shot thanks to the strong signal power, with quantum state tomography~\cite{Eichler2012}. For a normally open transistor, ideally one would expect the gate photon reflection in the vacuum state when the transistor is ``on'', or in the single-photon state when the transistor is switched off. As shown in the inset of Fig.~\ref{switch}\textbf{a}, in the experiment we measure a close-to-vacuum Wigner function for the reflected gate photons conditioned on the strong transmission, with an average gate photon number of about 0.03. On the other hand, conditioned on the weak transmission we obtain a Fock-state-like Wigner function with an average photon number of about 0.72. Those results demonstrate that the input signal is indeed switched by a single gate photon.

The reasons for the deviation of the measured conditional reflection of gate photons from the ideal vacuum state or Fock state are mainly twofold. Firstly the gate photon induced qubit flip process is not perfect. In the case of Fig.~\ref{switch}\textbf{a}, we measure a wrong qubit flip probability of $0.2$ for a single gate photon, and $0.04$ for zero gate photon. At the same time, the qubit may suffer from energy relaxation from the excited state or spurious qubit state transition during the input signal passing through the transistor, which leads to the non-zero probability of the ``off'' events when there is no gate photon, and thus introduces additional errors to the switching process. More detailed discussions can be found in Supplementary Note 5.

\bigskip
\textbf{Assume an ideal single-photon source.} To measure the switching probability of the transistor in the presence of a single gate photon, we separately calibrate the operation of the transistor in two stages, namely the qubit flip process and the following signal switch process. To calibrate the qubit flip probability by an incoming single gate photon, we send weak coherent state photons with varied average photon numbers to cavity I and measure the qubit flip rate. Ideally, the measured qubit flip rates shall follow the single-photon occupations of the weak coherent states, and thus from the experiment, we could determine the single-photon-induced qubit flip probability $\eta$ =0.80, which is mainly limited by the internal loss of cavity I (Supplementary Note 6).  To determine the qubit-state-flip-induced signal switch effect, we set the qubit to $\ket{e}$ and perform repeated single-shot measurements of the transmitted photon number. Qubit state relaxation to $\ket{g}$ would result in a wrong switch operation, and thus the measured transmission shows a bimodal distribution, from which we can determine the probability of correct switching induced by the qubit flip process $P_s = 0.925 $ for an input signal photon number of 37.2. The single-photon switching probability is given by $P_{sg}=\eta \times P_s = 0.74$. Details about the measurement can be found in the Methods section.

\bigskip
\textbf{Switch a strong signal.} 
A strong input signal can be effectively switched by our single-photon transistor. Fig.~\ref{switch}\textbf{c} shows the statistical working state of the transistor in repeated measurements with an average signal photon number $n_s=2.62\times10^5$, where the transistor is operated in normally open mode by aligning the signal frequency to the bare frequency of cavity II (see Supplementary Note 8 for details). Remarkably, for such a strong input signal one could still see a clear switch effect when the transistor is gated by weak coherent state photons, which agrees well with the theoretical predictions. It should be noted that when there is no gate photon, the wrong-operation events increase compared with the case with a weak input signal shown in Fig.~\ref{switch}\textbf{a}, which can be explained by the input signal induced qubit flip during the operation of the transistor~\cite{Siddiqi2012}. 

The single-photon switching character for the strong input signal is again verified through quantum state tomography on the reflected gate photons, as shown in the inset of Fig.~\ref{switch}\textbf{c}. We obtained an average number of reflected gate photons of $0.54$ conditioned on the weak transmitted signal, and 0.17 for the strong transmission. Compared with the results shown in Fig.~\ref{switch}\textbf{a}, the single-photon switching character shows a larger deviation from the ideal case, which is due to the stronger mixing of qubit states induced by the intense input signal~\cite{Siddiqi2012}.

\bigskip
\textbf{A normally closed transistor.} It is also possible to operate the transistor in a normally closed mode by controlling the qubit phase. As illustrated in Fig.~\ref{scheme}\textbf{d}, we set the phase of the second $\pi/2$ pulse to $\pi$, and thus the presence of a single gate photon would set the qubit state in $\ket{e}$ or $\ket{f}$, resulting in a strong transmission of signal photons from cavity II. 
Fig.~\ref{switch}\textbf{b} and \textbf{d} show the statistical transmission for a weak ($n_s=37.2$) and a strong ($n_s=2.62\times10^5$) input signal pulse, respectively.
In contrast to the case for a normally open transistor, the bimodal distribution remains, but with reversed weights for a strong output and a weak output. At the same time, the presence of gate photons introduces a clear reversal in the working status of the transistor. 
Together with the conditional state tomography on the reflected gate photons, those results confirm that here a single gate photon can effectively switch on the normally closed transistor.

\begin{figure}[!tbp]
\centering
\includegraphics[width=0.9\linewidth]{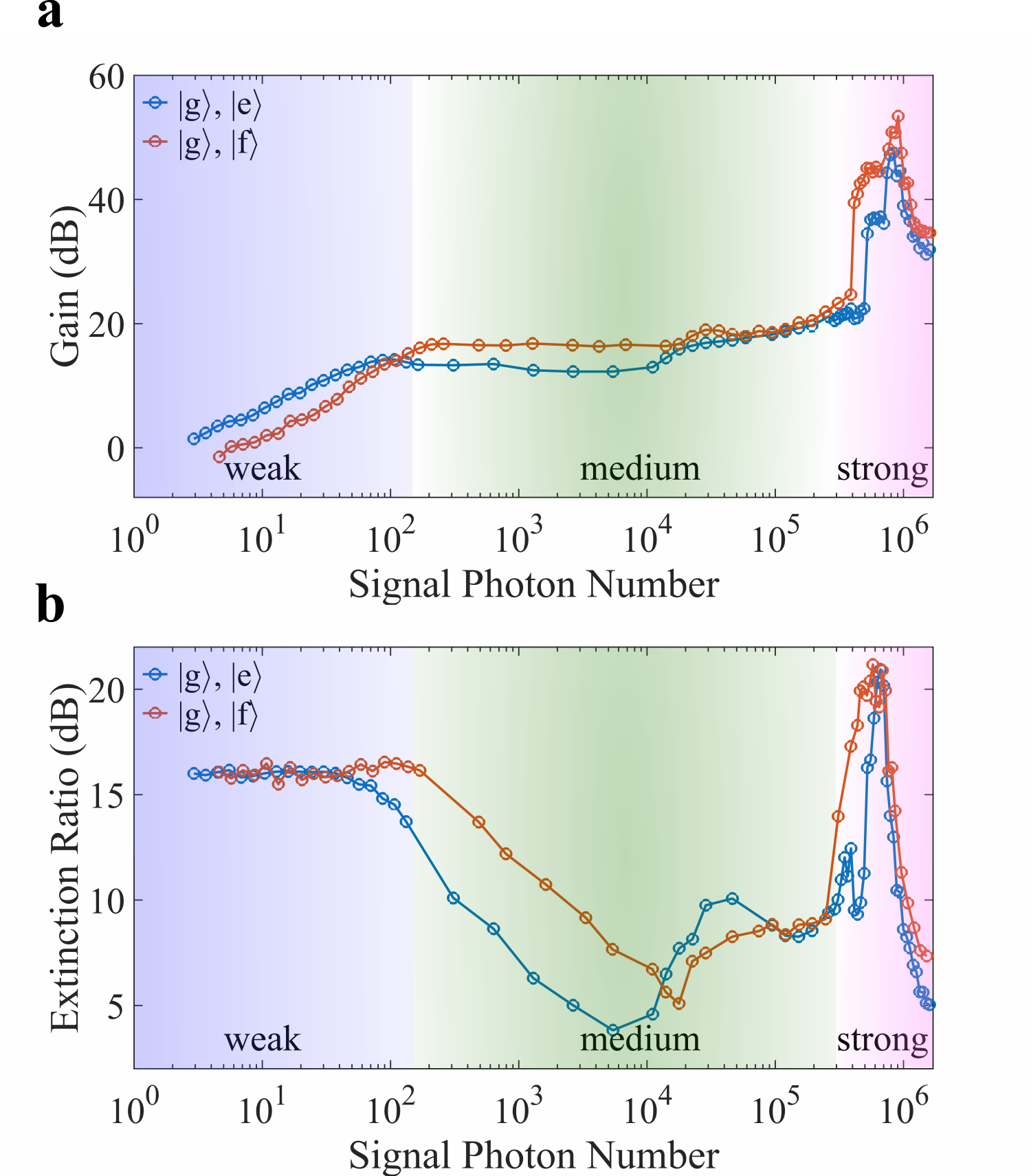}
\caption{\textbf{Gain and extinction ratio.} \textbf{a} and \textbf{b} show the extracted gain and extinction ratio of our device, respectively, when the transistor is fed with varied signal photon numbers from $3$ to $1.6\times 10^6$. According to the signal strength and the resulting performance of the transistor, three operation ranges of the transistor are distinct with different colors. Blue and red plots are the experimental results when the qubit is operated in the $\{\ket{g},\ket{e}\}$ subspace or $\{\ket{g},\ket{f}\}$ subspace, correspondingly. Note that since we pick up the best-measured gain and extinction ratio for each of the input signal strengths, the signal frequency is not necessarily kept unchanged.}

\label{gain}
\end{figure}

\bigskip
\textbf{Gain and extinction ratio.} As previously indicated, the measured transmission intensity and switching characters are largely determined by the number distribution of gate photons. To exclude the influence of gate photon source on the performance evaluation of the transistor, we compare the average transmission intensity with and without the gate photons, and extract the gain and extinction ratio based on the measurement results. Details of this derivation can be found in the Methods. The gain of a transistor describes the signal photon number that can be controlled by a single gate photon, which is defined as $G=10\log_{10}(|n^{open}_{\ket{1}}-n^{open}_{\ket{0}}|)$ for a normally open transistor, where $n^{open}_{\ket{1}}$ or $n^{open}_{\ket{0}}$ represents the average photon number outputted from the transistor gated either by a perfect single-photon source or a vacuum field, respectively. The extinction ratio is defined as $R=10\log_{10}(n^{open}_{\ket{1}}/n^{open}_{\ket{0}})$, which essentially describes the ``on''/``off'' contrast of the gated input.

Fig.~\ref{gain} shows the gain $G$ and the extinction ratio $R$ of our device as a function of the signal photon number, using either $\{\ket{g},\ket{e}\}$ subspace or $\{\ket{g},\ket{f}\}$ subspace. When the signal contains less than 100 photons, the transistor gain is proportional to the input photon number, which can exceed 16.6 dB when the qubit is operated in the $\{\ket{g},\ket{e}\}$ subspace, or 14.1 dB in the $\{\ket{g},\ket{f}\}$ subspace. The extinction ratio $R$ keeps above 15 dB when using both of the two qubit subspaces. In this regime, the extinction ratio is mainly determined by the cavity linewidth and dispersive shift. From the transmission spectra shown in Fig.~\ref{scheme}\textbf{e}, one would find an upper bound of the extinction ratio as large as 20 dB, which is substantially larger than the results shown in Fig.~\ref{gain}\textbf{a}. This is mainly caused by the non-ideal gating process and qubit decoherence during the operation of the transistor.

When the signal power increases, the gain of the transistor stops linear growth and keeps at around 16 dB. At the same time, the extinction ratio shows a drastic drop to even below 5 dB for some signal powers. This is due to the inherent non-linearity of the cavity mode when it is hybridized with the qubit state. The non-linearity is of minor importance for a weak input signal. When the photon number of the feeding signal exceeds 100, the photon population in the non-linear cavity would shift its resonant frequency, and thus prohibit the transmission of later incidence photons, which is known as the photon-blockade effect and explains the saturated gain and reduced extinction ratio in the medium range of the input signal power.

As reported in previous works~\cite{Blais2010,Girvin2010,Schoelkopf2010}, the cavity's non-linearity can be suppressed by populating the cavity with a large number of photons, which would effectively decouple the qubit and the cavity, and recover the linear response of the cavity to the feeding signal. Prominently, the critical photon number for this transition shows a strong dependence on the qubit state. More details about the transition can be found in Supplementary Note 8. This effect enables our device to reach an unprecedented level of gain. As shown in Fig.~\ref{gain}, when the signal photon number exceeds $4.9\times 10^5$, the gain of the transistor shows rapid increase to a peak value of 47.6 dB when operating the qubit in $\{\ket{g},\ket{e}\}$ subspace, and 53.4 dB in $\{\ket{g},\ket{f}\}$ subspace. The extinction ratio also increases from below 10 dB to about 20.5 dB peak value in this regime. 
For a further increased signal strength the cavity transmission shows minor dependence on the qubit state, and thus the presence of gate photons, resulting in a gradually failed switch operation.

\bigskip

\section{Discussion}
In summary, we have demonstrated a single-photon transistor with ultra-high gain in the microwave regime. Our device takes the advantage of the strong nonlinear interaction between a single microwave photon and a superconducting transmon qubit, which facilitates an effective photon-photon interaction at the single-photon level. We have shown that within a large power range the input signal fed to the transistor can be effectively switched by a single gate photon, with an unprecedented gain up to 53.4 dB, together with an extinction ratio better than 20.5 dB. We note that the cavity loss and qubit coherence bottleneck the single-photon switch probability for the current device. A further improvement in those aspects could improve the single-photon gating efficiency to 0.93. Moreover, benefiting from the non-demolition nature of the reflection-based photon-gating process, the operation of the transistor can be heralded by a single-photon detector if a low-efficient gate photon source is used. Our work paves the way for microwave-photon-based emerging technology for quantum information processing~\cite{Jonathan2020review,GU2017,Ritsch1997,qram2008}.



\section*{Methods} 
\textbf{The device.} The device is implemented with a 3D circuit quantum electrodynamics architecture~\cite{JC1,Wallraff2021}. The superconducting transmon qubit is patterned on a $1.2\,$mm$\times$15$\,$mm sapphire substrate with standard microfabrication techniques. The Josephson junction of the qubit is fabricated with electron-beam lithography and double-angle evaporation of aluminum. The qubit chip is placed in the waveguide trenches and dispersively coupled to the two 3D aluminum cavities. Detailed parameters of the device can be found in Supplementary Note 1.

\bigbreak
\textbf{Count ``on''/``off'' state of the transistor.} In the experiment, the ``on''/``off'' state of the transistor is determined by measuring the transmission intensity of the input signal. For a certain input signal strength, the transmitted photon number in each trial of the repeated measurements is recorded and summarized as a histogram plot as shown in Supplementary Figure 2. In general, the histogram shows a bimodal distribution, which corresponds to the ``on'' state and ``off'' state of the transistor. To classify the state of the transistor in each trial of the measurements, we use the K-means clustering method to determine the threshold of the histogram. It is worth mentioning that the width for each peak of the bimodal distribution is not determined by the performance of the transistor, but is mainly influenced by the detection efficiency of the amplification chain~\cite{Eichler2011,Lehnert2011,Eichler2012}. Therefore, we are only interested in the total count number for each transistor's working state. The extracted ``on''/``off'' state of the transistor is then summarized as bar plots in Fig.~2. We also calculated the average output photon number for either the ``on'' state or ``off'' state of the transistor based on the histogram plot, which is used as the x-axis position of the bar plots.

\bigbreak
\textbf{Calibrate the single-photon switching probability.} As mentioned in the main text, the operation of the single-photon transistor relies on gate-photon-induced qubit flip (cavity I) and the resulting signal switching (cavity II). To determine the single-photon switching probability, we separately measure the success probabilities of the two processes. It is known that in the small-photon-number regime, single-photon occupation in a weak coherent state is linearly related to the average photon number. Therefore we can directly measure the qubit flip probabilities with varied average gate photon numbers, and fit the experimental results to extract the qubit flip probability by a single gate photon $\eta$, as illustrated in Supplementary Figure 4. We develop a theoretical model for the single-photon-gating process, indicating that internal loss of cavity I during the photon-gating process is the main error source. A further improvement of the cavity performance and the qubit coherence would improve the gating efficiency to above 93\%. A detailed discussion can be found in Supplementary Note 6.

In another experiment, we directly set the qubit to $\ket{e}$ and perform repeated single-shot measurements on the transmitted signal photon number. Normally one could observe a bimodal distribution of the transmitted photon number, as a result of the unavoidable qubit relaxation to $\ket{g}$ when the signal passes through the transistor (see Supplementary Figure 2 as an example). We can count the ``on''/``off'' state from the bimodal histogram plot, and thus determine the success probability of the signal switching conditioned on the qubit flip $P_s$. The single-photon switching probability is thus given by $P_{sg}=\eta \times P_s$.

\bigbreak
\textbf{Extract gain and extinction ratio.} As discussed in the main text, when using the weak coherent state photons to gate the transistor, the measured switch effect is largely determined by the imperfection of the gate photon source. In order to obtain the intrinsic properties of the transistor, and accordingly extract the gain and extinction ratio of the transistor, we perform a series of controlled experiments to evaluate the switch effect by excluding the influence of the gate source. To be specific, we operate the transistor in either normally open (by setting $\theta=0$) or normally closed ($\theta=\pi$) mode. In both case we measure the transmission intensity when the transistor is either gated by the weak coherent state with $n_g=0.18$, or not gated. For the four cases, the measured transmission intensity can be written as
\begin{equation}
\begin{split}
n_{\ket{0}}^{open}&=P_g^{open} n_{\ket{g}}+(1-P_g^{open}) n_{\ket{e}}\\
n_{\ket{\alpha}}^{open}&=(P_g^{open}+\beta)n_{\ket{g}}+(1-P_g^{open}-\beta) n_{\ket{e}}\\
n_{\ket{0}}^{close}&=P_g^{close} n_{\ket{g}}+(1-P_g^{close}) n_{\ket{e}}\\
n_{\ket{\alpha}}^{close}&=(P_g^{close}-\beta)n_{\ket{g}}+(1-P_g^{close}+\beta) n_{\ket{e}}
\end{split}
\end{equation}
Here, $n_{\ket{0}(\ket{\alpha})}^{open}$ denotes the measured transmission intensity without gate photon (with weak coherent state gate photons) when the transistor is operated as a normally open mode. $n_{\ket{g}(\ket{e})}$ refers to the transmission intensity when the qubit is in $\ket{g}$ ($\ket{e}$) before the input signal is sent to the transistor. $P_g^{open(close)}$ denotes the probability of qubit state in $\ket{g}$ when the transistor is working as normally open (normally closed). $\beta$ is the qubit flip probability when the transistor is gated by the weak coherent state photons, which can be directly measured in the experiment (scattered plots in Supplementary Figure 4). For $n_g=0.18$ used in our experiment, we measure $\beta=0.13$. By solving these equations, $P_g^{open},P_g^{close},n_{\ket{g}},n_{\ket{e}}$ can be obtained.

Hereafter, We can directly calculate the expected transmission intensity if the transistor is gated by an ideal single photon source. As an example, the predicted transmission intensity when a normally open transistor is in the``on'' state ($n_{\ket{0}}^{open}$) or switched off by a single photon ($n_{\ket{1}}^{open}$) can be written as
\begin{equation}
\begin{split}
n_{\ket{0}}^{open}&=P_g^{open} n_{\ket{g}}+(1-P_g^{open}) n_{\ket{e}},\\
n_{\ket{1}}^{open}&=(P_g^{open}+\eta)n_{\ket{g}}+(1-P_g^{open}-\eta) n_{\ket{e}},
\end{split}
\end{equation}
where $\eta=0.80$ is the single-photon gating efficiency discussed in Supplementary Note 3. According to the definition mentioned in the main text, we can accordingly calculate the gain and extinction ratio based on the $n_{\ket{0}}^{open}$ and $n_{\ket{1}}^{open}$.

\bigbreak

\bigskip

\textbf{Data Availability:} The data generated in this study have been deposited in the zenodo database without accession code https://doi.org/10.5281/zenodo.7093109. Contact the corresponding author with any further questions.

\bigskip

\textbf{Code Availability:} The codes that support the findings of this study are available from the authors upon request.

\bigskip

\textbf{Acknowledgments:} This work was supported by Innovation Program for Quantum Science and Technology (2021ZD0301704), the Tsinghua University Initiative Scientific Research Program, the Ministry of Education of China, and the National Natural Science Foundation of China under Grant No.11874235 and No.11925404.

\bigskip

\textbf{Author Contributions:} L.M.D. and H.Y.Z. proposed and supervised the experiment. Z.H.B. and Y.L. prepared the sample with the assistance of W.Z.C., W.T.W., and L.Y.S. Z.L.W. and H.Y.Z. collected and analyzed the data with the assistance of Z.H.B., Y.L., X.Y.H., J.H.W., and Y.P.S. Z.L.W. carried out the theoretical analysis with the assistance of Y.K.W. Z.L.W., H.Y.Z. and L.M.D. wrote the manuscript.

\bigskip

\textbf{Competing interests:} The authors declare that there are no competing interests.

\bigskip

\textbf{Author Information:} Correspondence and requests for materials should be addressed to L.M.D. (lmduan@tsinghua.edu.cn) or H.Y.Z. (hyzhang2016@tsinghua.edu.cn).

\end{document}